# Determination of Molecules Distribution of the Impurity in Monocrystal of the Solid Solution by the Method of Raman Effect

M. A. Korshunov[1]

Kirensky Institute of Physics, Siberian Division, Russian Academy of Sciences, Krasnoyarsk, 660036 Russia

**Abstract** – distribution of centrosymmetrical molecules of an impurity (p-diclorobenzene) in monocrystals of solid solutions in two different matrixes with centrosymmetrical (p-dibrombenzene) and noncentrosymmetrical (p-bromchlorbenzene) molecules by the method of a Raman Effect is determined.

Organic crystals of a low symmetry find the increasing application in the molecular electronics engineering. Perspectives of use of such crystals, in particular of mixed ones, for an entry and handling of the information [1] are emphasized. Thus it is necessary that distribution of molecules of components in volume of the mixed crystal was the uniform. In order to achieve this it is necessary to have the knowledge of regularities of distribution of impurities on a crystal. It favors understanding of the mechanism of growth of crystals and enables to influence it. In contrast to metal alloys the components of organic mixed crystals may consist of as centrosymmetrical molecules, as and noncentrosymmetrical.

With this purpose distribution of molecules of an impurity formed by centrosymmetrical molecules (p-diclorobenzene) in volume of two mixed crystals, at first, with centrosymmetrical molecules of a p-dibrombenzene and, second, with noncentrosymmetrical molecules of a p-bromchlorbenzene was investigated. These mixed crystals were chosen because their components are isomorphic among themselves and form solid solutions at any concentrations of components. Monocrystals of solid solutions were grown by Bridgman's method.

---

[1] E-mail: mkor@iph.krasn.ru

The p-bromchlorbenzene, as a p-dibrombenzene and p-diclorobenzene ($\alpha$-modification), crystallizes in centrosymmetrical space group $P2_1/a$ with two molecules in a unit cell due to statistically unregulated distribution of molecules concerning parasubstitution haloids [2]. These crystals were explored by an X-ray diffraction method [2], methods of a Raman Effect [3] and a NQR [4]. The X-ray diffraction data on the mixed crystals investigated in paper is not presented in literature.

In the given paper the method of a Raman Effect which allows to judge on spectra of the lattice and intramolecular vibrationas character of a disposition of molecules of an impurity among molecules of the basic crystal is used. In particular, whether on spectra it is possible to spot the solid solution was formed as substitution, introductions or molecules of an impurity are focused on boundaries between blocks of a lattice of a monocrystal, or the solid solution is a mechanical mixture of substances.

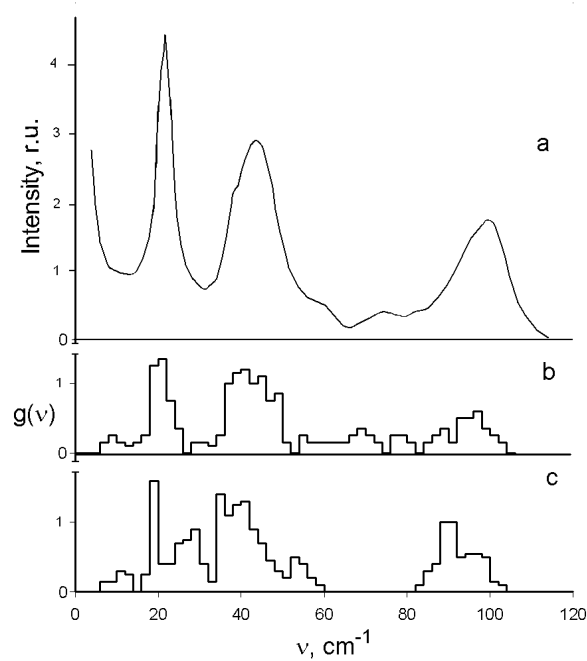

**Fig.1.** Experimental Raman spectrum (a) of the lattice vibrations of the mixed crystal of a p-dibrombenzene with p-diclorobenzene with components concentration of 50 mol. % and the histograms of spectra (b, c).

On fig. 1 the spectrum of the Raman effect (a) of the lattice vibrations of the mixed crystal of a p-dibrombenzene with p-diclorobenzene with equimolar concentration of 50 mol. % and the histograms of spectra (b, c). The histograms was calculated by Dyne's method [5] (histograms, obtained at calculations, show probability of appearing of spectrum lines in the chosen frequency interval). There two different suggestions presented in fig.1: the idea that the mixed crystals are formed by substitution (b) and idea that it is a mechanical mixture of components (c). From direct comparison of the experimental spectra (a) with histograms (b, c), obtained by Dyne's method, it is obvious, that these crystals are formed by substitution.

Magnitude of concentration of an impurity in the grown monocrystals of solid solutions was determined by the relative intensity of the valence intramolecular vibrations. According to paper [6], line in a spectrum of the Raman Effect of a p-dibrombenzene with frequency $v=212.0$ cm$^{-1}$ corresponds to valence vibration C-Br, and a line with frequency $v=327.0$ cm$^{-1}$ in p-diclorobenzene corresponds to valence vibration C-Cl. In fig. 2 the spectra of intramolecular vibrations (in area from 150 up to 400 cm$^{-1}$) of p-dibrombenzene (a), p-diclorobenzene (b) and p-bromchlorbenzene (d) and investigated mixed crystals with equimolecular concentrations of p-diclorobenzene in p-dibrombenzene (c) and in p-bromchlorbenzene (f) are presented.

Obviously, spectra of the mixed crystals are a superposition of spectra of intramolecular vibrations of components with taking into account their concentration. Using this method, concentrations of an impurity (p-diclorobenzene) in the mixed crystals which have been grown by the Bridgman's method were determined.

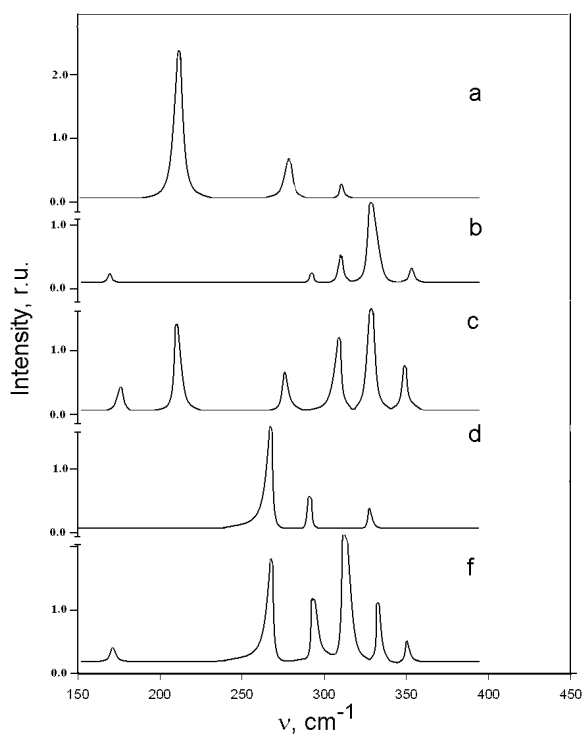

**Fig.2.** Experimental spectra of the intramolecular vibrations of a p-dibrombenzene (a), p-diclorobenzene (b) a p-bromchlorbenzene (d) and investigated mixed crystals with equimolecular concentrations of p-diclorobenzene in a p-dibrombenzene (c) and in a p-bromchlorbenzene (f).

The monocrystal was grown in a

glass tube in diameter d = 1cm and length L= 10 cm with a drawn capillary. In a tube initial substances in the necessary percentage were put. After that, the air was pumped out from the tube and the tube was soldered.

The tube with substance was moving with velocity $V=8.3 \cdot 10^{-6}$-$8.9 \cdot 10^{-6}$ cm/sec in crystallization furnace. The lapse rate of temperature of the furnace was set by various winding of a heating coil and made dT/dl=7.6-7.7 grad/sec.

The brought up monocrystal on length was cut on tablets in height no more than 0.1cm. Then from these tablets, with the purpose of investigation of radial distribution, from centre to edge in a plane of an edge a series of parallelepipeds with a size of an edge 0.1-0.2 cm was cut out. Monocrystallinity of this samples was checked with help of polarization microscope. For these samples the spectra of the Raman Effect were obtained. On these spectra by the method described above the concentration of an impurity of p-diclorobenzene in the obtained samples was determined and the modification of concentration of components, both on length, and on diameter of monocrystals is detected.

The obtained experimental data of investigation of distribution of an impurity along an axis of the grown monocrystals are presented in fig. 3 where axis of ordinates corresponds to concentration of an impurity C in a mol. % (p-diclorobenzene) and abscissa axis corresponds to length of a monocrystal – h=L / 10.

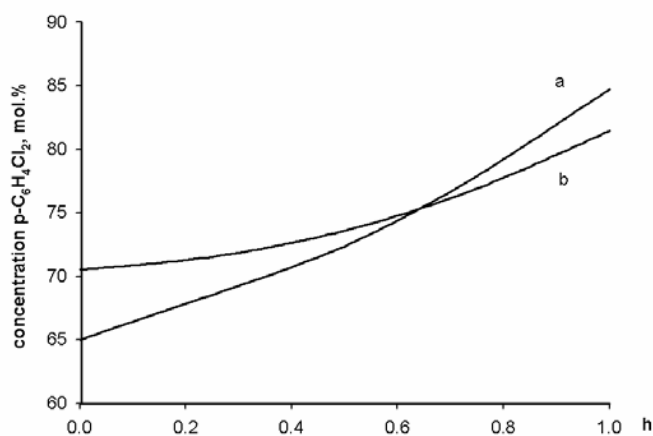

**Fig.3.** Distribution of impurity along an axis of the grown monocrystals.

Apparently, during growing of monocrystals the concentration of an impurity (p-diclorobenzene) was increasing. At the same time the change of concentration of p-diclorobenzene in a solid solution with p-bromchlorbenzene (graph. - a) is less, than in the mixed crystal with p-dibrombenzene (b).

In order to determine distribution coefficient k we will use following equation:

$$C_s = C_0 \cdot k \cdot (1-g)^{k-1},$$

where $C_s$ is the concentration of an impurity in the allocated portion of a solid state g, $C_0$ is the initial concentration of an impurity and k is a distribution coefficient which is the quantitative performance of distribution of an impurity between a solid state and a melt. From graphs on fig. 3 it is possible to determine distribution coefficients of an impurity of p-diclorobenzene in a p-dibrombenzene and in a p-bromchlorbenzene. The distribution coefficient of an impurity of p-diclorobenzene in a p-dibrombenzene is obtained and is equal to k=0.862, and in a p-bromchlorbenzene is equal to k=0.939.

If to consider a crystal of p-diclorobenzene as a solid solution with an impurity of molecules of p-diclorobenzene, than the graph of dependence of concentration C (for every $C_0$ =const) on length of a monocrystal h will represent a straight line parallel to an abscissa axis with a distribution coefficient k=1. As we can see, the distribution coefficient is declined from unity more for a solid solution of p-diclorobenzene/p-dibrombenzene than for p-diclorobenzene/ p-bromchlorbenzene.

Investigation of distribution of molecules of p-diclorobenzene in studied mixed crystals carried out with use of a method of the Raman effect along a plane (on diameter) that is perpendicular to the axis of a monocrystal, shows, that concentration of p-diclorobenzene will increase from centre to edge of a plane. Graphs on figs. 4, 5 shows distribution of molecules of p-diclorobenzene on diameter in various cuts of monocrystals (1-0.9h, 2 - 0.5h and 3 - 0.1h) solid solutions with a p-dibrombenzene (fig. 4) and a p-bromchlorbenzene (fig. 5).

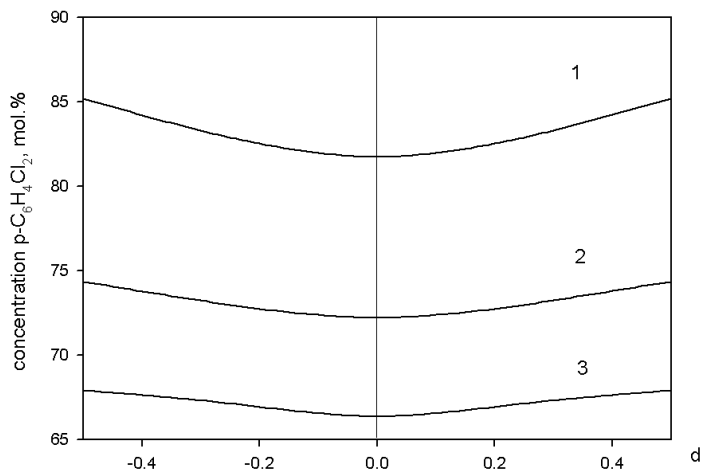

**Fig.4.** Distribution of molecules of an impurity (p-diclorobenzene) on diameter in various cuts of a monocrystal of a solid solution of a p-diclorobenzene / p-dibrombenzene (1 - 0.9h, 2 - 0.5h and 3 - 0.1h).

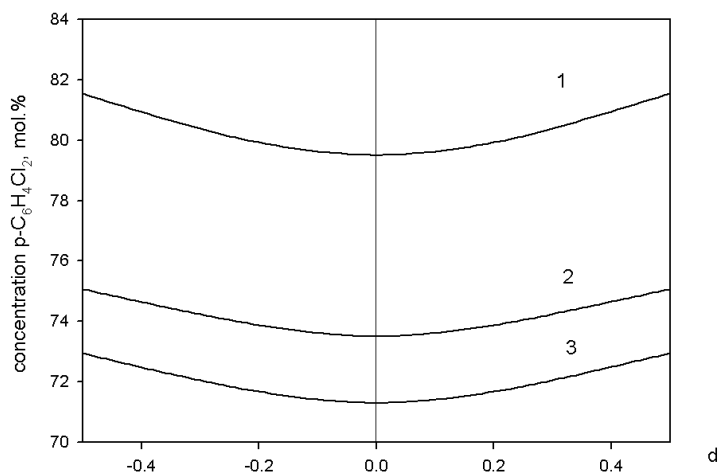

**Fig.5.** Distribution of molecules of an impurity (p-diclorobenzene) on diameter in various cuts of a monocrystal of a solid solution of a p-diclorobenzene / p-bromchlorbenzene (1 - 0.9h, 2 - 0.5h and 3 - 0.1h).

From these graph's data the distribution coefficients for various cuts of monocrystals were calculated. Results are presented in the table:

|       | 0.1h  | 0.5h  | 0.9h  |
|-------|-------|-------|-------|
| $K_1$ | 0.983 | 0.984 | 0.979 |
| $K_2$ | 0.988 | 0.989 | 0.988 |

Where $k_1$ is a distribution coefficient for a solid solution of p-diclorobenzene/p-dibrombenzene, and $k_2$ is a distribution coefficient for p-diclorobenzene/p-bromchlorbenzene.

Apparently from the table the deviation from unity is more in case of the mixed crystal of a p-diclorobenzene/p-dibrombenzene than for a p-diclorobenzene/p-bromchlorbenzene, as well as at study of distribution of an impurity along a monocrystal surveyed above.

Thus, the possibility of use the method of a Raman effect for investigation of distribution of molecules of components in monocrystals of the solid solutions which have been grown by the Bridgman's method is shown. Distribution of centrosymmetrical molecules of an impurity (p-diclorobenzene) in monocrystals of solid solutions in two different matrixes with centrosymmetrical (p-dibrombenzene) and noncentrosymmetrical (p-bromchlorbenzene) molecules, as along the grown monocrystal and as on its radius is defined. It is found, that in both cases the distribution coefficient in a solid solution of p-diclorobenzene/p-dibrombenzene is less than in a solid solution of p-diclorobenzene /p-bromchlorbenzene.

## References

1. New Physical Principles of Optical Data Processing, Akhmanov,S. A. and Vorontsov, M. A. Eds., Mocow: Nauka, 1990.
2. Kitaigorodskii, A. I., Rentgenostrukturnyi analiz, (X-ray Analysis), Moscow: Nauka, 1973.
3. Shabanov, V. F. and Korshunov, M. A., Fiz. Tverd. Tela (St. Petersburg), 1995, vol. 37, no. 11, pp. 3463-3469.
4. Grechishkin, V. S., Nuclear Quadrupole Interactions in Solids, Moscow: Nauka, 1971.
5. Dean, P., Numerical Methods in the Theory of Solids, Moscow, 1975, pp. 207-298.
6. Suzuki M., Ito M., Spectrochimica Acta. 1969. Vol. 25A. № 5 P. 1017.